\documentclass[useAMS,usenatbib,fleqn]{mnras}

\usepackage{graphicx,color}
\usepackage{mathtools}	%needed for dcase environment, loads amsmath package
\usepackage{amsmath}
\usepackage{amssymb}
\usepackage{bm}
\usepackage{subfig}             %for creating subfigures
\usepackage{ulem}
\usepackage{fixltx2e} %to display {figure*}'s in right order
\usepackage{pstricks} % PSTricks package
   \usepackage[british]{babel}             % British English hyphenation
   \usepackage{newtxtext}                  % Good fonts
   \usepackage[slantedGreek]{newtxmath}    %   "    "   (slanted Greek)
   \let\la=\lesssim % for less than similar from newtxmath, not \la from mnras.cls
   \usepackage[T1]{fontenc}                % Good font encoding
\def\apj{ApJ}

\def\mnras{MNRAS}
\def\aap{A\&A}
\def\apjl{ApJ}

\def\physrep{PhR}
\def\pre{PhRvE}
\def\prl{PhRvL}
\def\apjs{ApJS}

\def\nat{Nature}

\def\ssr{SSRv}

\def\gapfd{GApFD}
\def\an{AN}

\renewcommand{\vec}[1]{{{\mbox{\boldmath $#1$}}}}%also makes bold Greek letters

\newcommand{\zhat}{\hat{z}}

\newcommand{\del}{\partial}
\newcommand{\Del}{{\nabla}}

\newcommand{\bmDel}{\bm{\nabla}}

\newcommand{\bmomega}{\vec{\omega}}
\newcommand{\bfOmega}{\vec{\Omega}}

\newcommand{\bfgamma}{\vec{\gamma}}

\newcommand{\Emf}{\bm{\mathcal{E}}}

\newcommand{\bfB}{\bm{B}}
\newcommand{\bfJ}{\bm{J}}
\newcommand{\bfu}{\bm{u}}
\newcommand{\bfb}{\bm{b}}

\newcommand{\Emfr}{\mathcal{E}_r}
\newcommand{\Emfp}{\mathcal{E}_\phi}
\newcommand{\Emfz}{\mathcal{E}_z}

\newcommand{\mean}[1]{\overline{#1}}
\newcommand{\meanv}[1]{\bm{#1}}

\newcommand{\f}{_\mathrm{0}}					   	%forcing
\newcommand{\forc}{_\mathrm{f}}					   	%forcing
\newcommand{\kin}{_\mathrm{k}}			   		%kinematic
\newcommand{\magn}{_\mathrm{m}}			   		%magnetic
\newcommand{\crit}{_\mathrm{c}}			   		%critical
\newcommand{\diff}{_\mathrm{d}}			   		%diffusion, subscript
\newcommand{\cro}{\times}
\newcommand{\Rm}{\mathcal{R}_\mathrm{m}}
\newcommand{\Rey}{\mathcal{R}_\mathrm{e}}
\newcommand{\mbr}{B_r}
\newcommand{\mbp}{B_\phi}
\newcommand{\mbz}{B_z}

\newcommand{\muz}{U_z}

\newcommand{\alphatilde}{\widetilde{\alpha}}

\newcommand{\Dtilde}{\widetilde{D}}

\renewcommand{\theenumi}{\roman{enumi}}

\newcommand{\Pm}{\mathrm{Pm}}

  \newcommand{\kms}{\,{\rm km\,s^{-1}}}

  \newcommand{\kpc}{\,{\rm kpc}}
  \newcommand{\pc}{\,{\rm pc}}
  
  \newcommand{\Myr}{\,{\rm Myr}}

  \newcommand{\nG}{\,{\rm nG}}

\title[A new constraint on galactic dynamo theory]
{A new constraint on mean-field galactic dynamo theory}
\author[L.\ Chamandy, N.\ Singh]{Luke Chamandy,$^{1,2,3}$\thanks{lchamandy@pas.rochester.edu}
\& Nishant K. Singh$^{4,5}$\thanks{singh@mps.mpg.de}\\
$^{1}$Department of Physics and Astronomy, University of Rochester, Rochester NY, 14618, USA\\
$^{2}$Astronomy Department, University of Cape Town, Rondebosch 7701, Republic of South Africa\\
$^{3}$Department of Physics, University of the Western Cape, Belleville 7535, Republic of South Africa\\
$^{4}$Nordita, KTH Royal Institute of Technology and Stockholm University, Roslagstullsbacken 23,
SE-10691 Stockholm, Sweden\\
$^{5}$Max Planck Institute for Solar System Research,
Justus-von-Liebig-Weg 3, D-37077 G\"ottingen, Germany
}

\begin{document}

\pagerange{\pageref{firstpage}--\pageref{lastpage}} \pubyear{2016}

\maketitle

\begin{abstract}
Appealing to an analytical result from mean-field theory,
we show, using a generic galaxy model,
that galactic dynamo action 
can be suppressed by small-scale magnetic fluctuations.
This is caused by the magnetic analogue 
of the R\"{a}dler or $\bfOmega\cro\bfJ$ effect,
where rotation-induced corrections to the
mean-field turbulent transport
result in what we interpret to be
an effective reduction of the standard $\alpha$ effect
in the presence of small-scale magnetic fields.
\end{abstract}

\label{firstpage}
\defcitealias{Brandenburg+Subramanian05a}{BS05}
\defcitealias{Radler+03}{RKR}
\defcitealias{Chamandy+14b}{CSSS}
\defcitealias{Chamandy+Taylor15}{CT}
\defcitealias{Squire+Bhattacharjee15c}{SB15}
%----------------------------------------------------------------------------
\section{Introduction}
\label{sec:intro}
Astrophysical dynamos can be loosely divided into small-scale
(or fluctuation) dynamos, 
which amplify the field on scales up to the outer scale,
and large-scale (or mean-field) dynamos, 
which amplify the field on larger scales up to the system size.
Both are expected to occur simultaneously 
in turbulent astronomical bodies such as stars and galaxies.
There have been some attempts to understand these in a
single unified framework
\citep{Subramanian99,Subramanian+Brandenburg14,Bhat+16}.
Nevertheless, given sufficient scale-separation,
it seems reasonable to treat them as distinct, 
yet interconnected, entities \citep{Brandenburg+Subramanian05a,Brandenburg+12a}.

The small-scale dynamo operates much faster than the large-scale dynamo.
In galaxies, the former 
exponentiates the field on a timescale of order
the shortest eddy turnover time,
whilst the latter has an e-folding time 
that is probably limited from below by the 
galactic rotation period.
As the small-scale magnetic field saturates
near energy equipartition with turbulence,
it could significantly influence already growing large-scale magnetic
field by modifying mean-field transport coefficients
\citep[e.g.][]{Radler+03,Brandenburg+Subramanian05a}
(hereafter \citetalias{Radler+03}; 
\citetalias{Brandenburg+Subramanian05a}).
This is expected to affect not only the growth/decay rates, but
also the saturation level of the large-scale magnetic field.
Importantly, such effects are often proportional 
to the second moment of the small-scale magnetic field,
so relevant even for non-helical magnetic field.

Part of the reason such effects 
have been mostly ignored in models
may be that they tend to be associated with anisotropy in the 
turbulent transport coefficients,
which is often neglected for simplicity 
\citep[but see e.g.][]{Gressel+08b,Pipin+Seehafer09}.
On the other hand, 
numerical studies that have calculated the
transport coefficients using the test field method 
\citep{Schrinner+05,Schrinner+07,Rheinhardt+Brandenburg10}
have so far been restricted to the regime where 
large- and small-scale magnetic field components are weak
\citep{Sur+08,Gressel+08a,Gressel+08b,Brandenburg+08a,Brandenburg+12b}.

A different numerical approach was taken by \citet{Squire+Bhattacharjee15b}.
They concluded that the 
shear-current effect
\citep{Rogachevskii+Kleeorin03,Rogachevskii+Kleeorin04}
can drive dynamo action in the presence of magnetic fluctuations,
for moderate values of the magnetic Reynolds number $\Rm$.
Likewise, a magnetic contribution to the
$\bfOmega\cro\bfJ$ 
effect \citep{Radler69} 
was found by \citet{Squire+Bhattacharjee15b} to play a role.
Their results were supported 
with analytical calculations using quasilinear theory
\citep{Squire+Bhattacharjee15c}.
However, 
it is not clear what the implications are for the
large $\Rm$ and fluid Reynolds number $\Rey$ 
(along with Strouhal number $\sim1$)
regime relevant for galaxies.

In the present work, we apply the basically equivalent 
results of \citetalias{Radler+03,Brandenburg+Subramanian05a},
who calculated the mean electromotive force (emf)
for turbulence with slow rotation and weak stratification,
to a simple mean-field galactic dynamo model.
We show that a magnetic R\"{a}dler effect 
competes with,
and partially suppresses, the $\alpha$ effect 
responsible for the generation of poloidal mean-field from toroidal.

\section{Mean emf}
\label{sec:slowrotation}
The mean induction equation is
\begin{equation}
  \label{dynamo}
  \del\meanv{B}/\del t= \bmDel\cro\left( \meanv{U}\cro\meanv{B} +\Emf\right),
\end{equation}
where $\Emf=\mean{\bfu\cro\bfb}$ 
is the mean emf and Ohmic terms have been neglected.
Here bar represents mean,
and we use uppercase (lowercase) 
to designate large-scale (small-scale) fields.
We adopt the expression for $\Emf$ 
from Sect.~10.3 of \citetalias{Brandenburg+Subramanian05a}, 
where helicity is induced by slow rotation  
and weak stratification of the turbulence,
and $\rho$ is assumed to be constant 
(the incompressible regime).
Large-scale shear is neglected.
The mean emf can then be expanded as
\begin{equation}
  \label{Emf}
   \meanv{\Emf}_i=  \alpha_{ij}B_j 
                   -\eta_{ij}J_j 
                   +(\bm{\gamma}\cro\meanv{B})_i 
                   +(\bm{\delta}\cro\meanv{J})_i 
                   +\kappa_{ijk}B_{j,k},
\end{equation}
where comma denotes partial differentiation,
and $\mu\f\meanv{J}=\bmDel\cro\meanv{B}$. 
Henceforth we adopt units such that $\rho=1$ and $\mu\f=1$
\citepalias[for details see][]{Radler+03,Brandenburg+Subramanian05a}.
\citetalias{Brandenburg+Subramanian05a} find
\begin{align}
  \label{alpha}
  &\alpha_{ij}= \tfrac{1}{3}\tau\delta_{ij}\mean{\bm{j}\cdot\bm{b}} \nonumber
    -\tfrac{4}{5}\tau^2\Big[\delta_{ij}\bmomega\cdot\bmDel
      \left(u^2-\tfrac{1}{3}b^2\right)\\
    &\qquad-\tfrac{11}{24}\left(\omega_i\Del_j 
      +\omega_j\Del_i\right)\left(u^2 +\tfrac{3}{11}b^2\right)\Big],\\
  \label{eta}
  &\eta_{ij}= \tfrac{1}{3}\tau\delta_{ij}u^2,\\
  \label{gamma}
  &\bm{\gamma}= -\tfrac{1}{6}\tau\bmDel\left(u^2 -b^2\right) 
     -\tfrac{1}{6}\tau^2\bmomega\cro\bmDel\left(u^2 +b^2\right),\\
  \label{delta}
  &\bm{\delta}= \tfrac{1}{6}\tau^2\bmomega\left(u^2 -b^2\right),\\
  \label{kappa}
  &\kappa_{ijk}= \tfrac{1}{6}\tau^2\left(\omega_j\delta_{ik} 
     +\omega_k\delta_{ij}\right)\left(u^2 +\tfrac{7}{5}b^2\right),
\end{align}
where $u=\sqrt{\mean{u^2}}$ is the rms turbulent velocity, 
$b=\sqrt{\mean{b^2}}$ is the rms small-scale magnetic field (or Alfven speed in our units), 
and $\bm{\omega}$ is the angular velocity.
One effect of rotation is to induce anisotropy of the turbulence
through the action of the Coriolis force.
Strictly speaking $u$ characterizes a presumed initial turbulent state
with low magnetic fields.
There is no such restriction on $b$ \citepalias{Brandenburg+Subramanian05a}.
Note that $\eta_{ij}$, $\bm{\delta}$, 
and $\kappa_{ijk}$ are independent of stratification, where
$\eta_{ij}$ describes the isotropic turbulent diffusion,
the $\bm{\delta}$-term
is often also known as the $\bm{\Omega \times J}$ or R\"adler
effect \citep{Radler69}, and the tensorial $\kappa$-term
involves a nontrivial diffusion of the mean magnetic field. 
Both $\delta$ and $\kappa$
effects, exhibiting a generalized diffusion,
lead to a cross-coupling of different components of the mean magnetic
field and thus share this property 
with the standard $\alpha$ effect.

Expressions~\eqref{alpha}--\eqref{kappa} were derived 
using the minimal $\tau$ approximation (MTA) \citep{Blackman+Field02}
with relaxation time $\tau$ assumed to be scale-independent,
but we consider the effects of relaxing this assumption 
(the spectral $\tau$ approximation) in Sect.~\ref{sec:spectraltau}.
For simplicity we solve the mean-field dynamo equations
under the quasilinear approximation,
which, unlike the MTA, neglects non-locality in time.
This is not expected to make a difference
when the e-folding time of mean-field dynamo growth or decay 
is much larger than the relaxation time $\tau$ \citep[e.g.][]{Chamandy+13b}.%
\footnote{To maintain consistency with results of the quasilinear approximation,  
$\tau$ is interpreted to be equal to the turbulence correlation time 
\citep[see also][]{Brandenburg+Subramanian05b}.}

We adopt cylindrical coordinates $(r,\phi,z)$ with $\bmomega=\Omega\hat{z}$
and apply the slab approximation suitable for a thin disk, 
which renders the problem 1D in $z$.
That is, we ignore all derivatives except $\del/\del z$,
with the exception of the radial derivative of $\Omega$,
which leads to the $\Omega$ effect
(which depends on the shear).
All turbulent transport coefficients then reduce to scalars,
and we obtain%
\footnote{The $\phi$-component of equation~\eqref{dynamo}
has a part $-\del\Emfz/\del r$
which contains a term proportional to $(\del\Omega/\del r)\mbz$.
The ratio of the magnitude of this contribution to that of  
the $\Omega$ effect is $\sim (h/r)(\tau^2u^2/h^2)\mbz/\mbr\ll1$,
where $h$ is the scale height.
Therefore we need not consider $\Emfz$ in what follows.}
\begin{align}
  \label{Emfr_slab}
  \Emfr&= \alpha\mbr +\eta\del\mbp/\del z -\gamma\mbp -(\delta-\kappa)\del\mbr/\del z,\\
  \label{Emfp_slab}
  \Emfp&= \alpha\mbp -\eta\del\mbr/\del z +\gamma\mbr -(\delta-\kappa)\del\mbp/\del z,
\end{align}
with
\begin{align}
  \label{alpha_slab}
  &\alpha= \alpha_{rr}= 
     \alpha_{\phi\phi}= \tfrac{1}{3}\tau\mean{\bm{j}\cdot\bm{b}} 
     -\tfrac{4}{5}\Omega\tau^2\frac{\del}{\del z}\left(u^2 -\tfrac{1}{3}b^2\right),\\
  \label{eta_slab}
  &\eta= \eta_{rr}= \eta_{\phi\phi}= \tfrac{1}{3}\tau u^2,\\
  \label{gamma_slab}
  &\gamma= \gamma_z= -\tfrac{1}{6}\tau\frac{\del}{\del z}\left(u^2 -b^2\right),\\
  \label{delta_slab}
  &\delta=\delta_z= \tfrac{1}{6}\Omega\tau^2\left(u^2 -b^2\right),\\
  \label{kappa_slab}
  &\kappa= \kappa_{rrz}= \kappa_{\phi\phi z}
     = \tfrac{1}{6}\Omega\tau^2\left(u^2 +\tfrac{7}{5}b^2\right).
\end{align}
Combining equations \eqref{delta_slab} and \eqref{kappa_slab} we find
\begin{equation}
  \label{deltakappa_slab}
  \delta'= \delta -\kappa = -\tfrac{2}{5}\Omega\tau^2b^2;
\end{equation}
that is, the $\delta$ and $\kappa$ effects 
reduce to one effect in the regime considered
\citep{Brandenburg+08a}, so below we use $\delta'$ for ease of notation.
Note that the component of $\delta'$ proportional to $u^2$
cancels out, 
leading to effects that depend on $b^2$ but not on $u^2$, 
and therefore we term it as the magnetic R\"adler effect.
In the slab approximation,
\begin{align}
  \label{curlEmf_r_slab}
  \left(\bmDel\cro\meanv{\Emf}\right)_r=  \;
    &-\frac{\del\Emfp}{\del z}= 
     -\frac{\del\alpha}{\del z}\mbp -\frac{\del\gamma}{\del z}\mbr 
     -\left(\alpha -\frac{\del\delta'}{\del z}\right)\frac{\del\mbp}{\del z}\nonumber\\ 
    &+\left(\frac{\del\eta}{\del z} -\gamma\right)\frac{\del\mbr}{\del z} 
     +\delta'\frac{\del^2\mbp}{\del z^2} +\eta\frac{\del^2\mbr}{\del z^2},\\
  \label{curlEmf_p_slab}
  \left(\bmDel\cro\meanv{\Emf}\right)_\phi=
    & \frac{\del\Emfr}{\del z}= 
      \frac{\del\alpha}{\del z}\mbr -\frac{\del\gamma}{\del z}\mbp
     +\left(\alpha -\frac{\del\delta'}{\del z}\right)\frac{\del\mbr}{\del z} \nonumber\\ 
    &+\left(\frac{\del\eta}{\del z} -\gamma\right)\frac{\del\mbp}{\del z}
     -\delta'\frac{\del^2\mbr}{\del z^2} +\eta\frac{\del^2\mbp}{\del z^2}.
\end{align}
Expressions \eqref{alpha_slab}--\eqref{deltakappa_slab} 
can now be substituted into equations~\eqref{curlEmf_r_slab}--\eqref{curlEmf_p_slab}. 
For simplicity we assume $\tau$ to be independent of $z$,
and we also assume $\del\Omega/\del z=0$, 
which is reasonable for a thin galactic disk.
The diamagnetic pumping term $\bfgamma\cro\meanv{B}$ in equation~\eqref{Emf}
is of the same form as the term in equation~\eqref{dynamo}
involving the vertical mean velocity $\muz \zhat$, 
and may lead to an effective reduction of $\muz$ \citep[e.g.][]{Gressel+08a};
here we make the simplifying assumption that $\gamma=0$.
Likewise we take $\del\eta/\del z=0$.
Finally, we neglect the terms proportional to $\del\delta'/\del z$
for reasons explained in Sect.~\ref{sec:alpha_k} below,
so that equation~\eqref{dynamo} reduces to
\begin{equation}
  \label{Br}
  \frac{\del B_r   }{\del t}=              -\frac{\del}{\del z}(\alpha B_\phi) 
                             +\delta'\frac{\del^2B_\phi}{\del z^2} +\eta\frac{\del^2B_r   }{\del z^2},
\end{equation}
\begin{equation}
  \label{Bp}
  \frac{\del B_\phi}{\del t}= -q\Omega B_r +\frac{\del}{\del z}(\alpha B_r   ) 
                             -\delta'\frac{\del^2B_r   }{\del z^2} +\eta\frac{\del^2B_\phi}{\del z^2},
\end{equation}
with $\mbz$ equal to a constant since solenoidality requires that $\del\mbz/\del z=0$.
Terms involving $\mbz$ are small, so neglected \citep{Ruzmaikin+88}.
Here $q= -\mathrm{d}\ln\Omega/\mathrm{d}\ln r$ is the local shear parameter,
equal to unity for a flat rotation curve.

\section{Effects of magnetic fluctuations}

\label{sec:alpha}
Below, we make use of both analytic and numerical solutions
of equation~\eqref{dynamo} for a slab geometry
and with vacuum boundary conditions 
at the disc surfaces.
The analytic solution neglects the $\alpha$ 
and $\delta'$ terms in the $\phi$-component, 
and also makes use of the `no-$z$' approximation
\citep{Subramanian+Mestel93,Phillips01}.
The numerical method solves the full 
equations~\eqref{Br} and \eqref{Bp}
for $\mbr(z,t)$ and $\mbp(z,t)$
with $\alpha\propto\sin(\piup z/h)$
and other parameters being constants.
Details of the galactic dynamo model
as well as analytical and numerical methods 
can be found in \citet{Chamandy+14b}
\citepalias[hereafter][]{Chamandy+14b}
and \citet{Chamandy+Taylor15}.
Simulations are already well-converged 
at the low resolution of $51$ grid points.

\subsection{$\alpha$ effect and $(\del\delta'/\del z)\del\mbp/\del z$ term}
\label{sec:alpha_k}
Defining $\alpha\kin$ to be the part of $\alpha$ that is not explicitly dependent on $b$
we find from equation~\eqref{alpha_slab},
\begin{equation}
  \label{alpha_kin_estimate}
  \alpha\kin= -\tfrac{4}{5}\Omega\tau^2\del u^2/\del z 
            \sim\Omega\tau^2u^2/h,
\end{equation}
with scale height $h$.
This is the standard estimate 
(\citealt{Krause+Radler80}, Ch.~VI of \citealt{Ruzmaikin+88}).
Invoking the `no-$z$' approximation we obtain
\begin{equation}
  \label{alpha_kin_noz}
  \left(\bmDel\cro\meanv{\Emf}\right)_r= 
    \ldots -\frac{\del}{\del z}(\alpha\kin\mbp) 
    \sim\ldots -\frac{2\Omega\tau^2u^2}{\piup h^2}\mbp,
\end{equation}
where dots represent other terms in the equation.

Now consider how $\alpha$ is affected by terms involving $b^2$,
excluding the current helicity term $\alpha\magn$. 
The latter term governs the so-called dynamical quenching 
\citep{Blackman+Field02,Shukurov+06}.
Substituting equations~\eqref{alpha_slab}
and \eqref{deltakappa_slab} into
equation~\eqref{curlEmf_r_slab},
we obtain
\begin{equation}
  \label{alpha_terms}
  \begin{split}
    \left(\bmDel\cro\meanv{\Emf}\right)_r= \ldots +\tfrac{4}{5}\Omega\tau^2
    &\left[\frac{\del^2}{\del z^2}(u^2 -\tfrac{1}{3}b^2)\mbp\right.\\
    &\;\,\left.+\frac{\del}{\del z}(u^2-\tfrac{5}{6}b^2)\frac{\del\mbp}{\del z}\right].
  \end{split}
\end{equation}
The coefficients of $b^2$ for the two terms are different ($-1/3$ vs. $-5/6$)
because of the contribution of the term proportional to $\del\delta'/\del z$.

In galactic dynamo models such as \citetalias{Chamandy+14b} 
the two terms in equation~\eqref{alpha_terms} have opposite sign,
and the first term must out-compete the second to obtain growing solutions.
Both terms in equation~\eqref{alpha_terms} get reduced when $b^2\sim u^2$. 
However, the second term is reduced more,
which helps the first term to win.
This would tend to effectively enhance the $\alpha$ effect
in certain galactic dynamo models,
but the overall effect is model-dependent,
since it depends on, e.g., the stratification of $u^2$.
In any case, the role of magnetic fluctuations 
\textit{may} not be very important here
given that in the first (dominant) term 
$b^2$ is multiplied by the factor $1/3$.
Hence we do not consider these contributions further,
though they may be important in some cases,
and we intend to explore them in detail in a subsequent work.

\subsection{$\delta'(\del^2\mbp/\del z^2)$ term}
\label{sec:delta_kappa}
More interesting (and hence the focus of 
the present work)
is the contribution of the small-scale magnetic field 
to the term in equation~\eqref{curlEmf_r_slab}
involving $\delta'$ and proportional to $\del^2\mbp/\del z^2$.
(This term has the same form as 
the term involving $\eta_{\phi r}\del^2\mbp/\del z^2$
would have had
if $\eta_{\phi r}$ had been finite.)
Now to gain insight into the effect of this term,
assume, for the moment, 
that the vertical profile (but not necessarily amplitude) 
of the mean magnetic field in the linear regime 
is comparable to that of standard galactic dynamo solutions.
Then the `no-$z$' approximation can be used
to estimate the contribution of this term.
We obtain
\begin{equation}
  \label{deltakappa_estimate}
  \begin{split}
    &\left(\bmDel\cro\meanv{\Emf}\right)_r=    
      \ldots +\delta'\frac{\del^2\mbp}{\del z^2}
      =\ldots -\tfrac{2}{5}\Omega\tau^2b^2\frac{\del^2\mbp}{\del z^2}\\
    &\quad\sim \ldots -\tfrac{2}{5}\Omega\tau^2b^2\left(-\frac{\piup^2\mbp}{4h^2}\right)
      =\ldots +\frac{\piup^2\Omega\tau^2b^2}{10h^2}\mbp,
  \end{split}
\end{equation}
which can be compared with the term involving $\alpha\kin$ in
equation~\eqref{alpha_kin_noz}.
The contributions
of equations~\eqref{alpha_kin_noz} and \eqref{deltakappa_estimate}
do not have the same form in $z$, in general.
However, they do have the same form under the `no-$z$' approximation.
We see then that for $b^2\sim u^2$, the above term involving $\delta'$
can suppress, or even cause an effective sign reversal of, $\alpha$.
Results reported below show that this `no-$z$' prediction
is borne out in numerical solutions.

\begin{table}
  \begin{center}
  \caption{Key dimensionless parameters for numerical examples.
           \label{tab:models}
          }
  \begin{tabular}{@{}lcccccc@{}}
\hline
Model     &$\xi$    &$R_\Omega$  &$R_\alpha$ &$q^{1/2}h/(\tau u)$           &$\Omega\tau$  &$D$           \\
\hline                                                                                                    
A$_\xi$   &varied   &$-14.1$     &$0.92$     &$3.91$                        &$0.31$        &$-13.0$       \\
A$\f$     &$0$      &varied      &$0.92$     &$1.04\sqrt{-R_\Omega}$        &$0.31$        &$0.92R_\Omega$\\
A$_{0.4}$ &$0.4$    &varied      &$0.92$     &$1.04\sqrt{-R_\Omega}$        &$0.31$        &$0.92R_\Omega$\\
B$_\xi$   &varied   &$-21.1$     &$1.38$     &$3.91$                        &$0.46$        &$-29.2$       \\
B$\f$     &$0$      &varied      &$1.38$     &$0.85\sqrt{-R_\Omega}$        &$0.46$        &$1.38R_\Omega$\\
B$_{0.4}$ &$0.4$    &varied      &$1.38$     &$0.85\sqrt{-R_\Omega}$        &$0.46$        &$1.38R_\Omega$\\
\hline
  \end{tabular}
  \end{center}
\end{table}

\subsection{When is the effect important?}
\label{sec:xi_crit}
It is convenient to define the parameter $\xi= b^2/u^2$,
the ratio of mean small-scale magnetic 
and kinetic energy densities.
Let us also define an effective $\alpha$, 
called $\alphatilde$, such that 
\begin{equation}
  \label{alphatilde_definition}
  -\frac{\del}{\del z}(\alphatilde\mbp)= -\frac{\del}{\del z}(\alpha\kin\mbp) 
                                 +\delta'\frac{\del^2\mbp}{\del z^2}.
\end{equation} 
Solving for $\alphatilde$,
using relation~\eqref{alphatilde_definition}
along with estimates~\eqref{alpha_kin_noz} and \eqref{deltakappa_estimate}
(with $\del/\del z\simeq2/(\piup h)$),
we obtain
\begin{equation}
  \label{alphatilde}
  \alphatilde\sim \alpha\kin +\frac{\piup^3}{8h}\delta'
             \sim \frac{\tau^2u^2\Omega}{h}\left(1-\frac{\piup^3}{20}\xi\right).
\end{equation}
This gives an estimate of the threshold value of $\xi$, 
$\xi\f\approx 20/\piup^3=0.65$, 
such that $\alphatilde=0$ if $\xi=\xi\f$,
and for $\xi>\xi\f$, 
the sign of $\alphatilde$ is opposite to that of $\alpha\kin$.
At this point it is useful to define 
the dimensionless Reynolds numbers:
$R_\Omega= -q\Omega h^2/\eta$ and $R_\alpha= \alpha\kin h/\eta$.
In the $\alpha\Omega$ limit of the $\alpha^2\Omega$ dynamo
with a purely toroidal mean velocity field $U=(0,r\Omega,0)$,
the kinematic growth rate depends only on the dynamo number
$D= R_\Omega R_\alpha$.

Let us now estimate the threshold value $\xi\crit<\xi\f$
for obtaining a supercritical dynamo.
Let us assume, for simplicity, 
that vertical outflows are too small to affect the dynamo.
The value we obtain for $\xi\crit$ is then an upper limit 
because such outflows generally weaken 
mean-field dynamo action 
in the linear (in 
$\meanv{B}$) regime.

In the $\alpha\Omega$ approximation,
the condition for a supercritical dynamo is $|D|>|D\crit|$,
where $D\crit$ can be determined numerically or estimated analytically.
The analytic solution \citepalias{Chamandy+14b}
gives $D\crit\approx -(\piup/2)^5$.
Defining $\Dtilde$ in a way analogous to $\alphatilde$ 
in equation~\eqref{alphatilde},
we get $\Dtilde= D(1-\xi/\xi\f)$.
Then $\xi\crit$ is obtained 
(under the `no-$z$' and $\alphatilde\Omega$ approximations)
by setting $\Dtilde=\Dtilde\crit=D\crit$
and solving for $\xi$,
which gives
\begin{equation}
  \label{crit}
  \xi\crit\sim \xi\f(1-D\crit/D) 
    \sim\frac{20}{\piup^3}\left(1-\frac{\piup^5u^2}{288qh^2\Omega^2}\right).
\end{equation}
Here we have made use of equations~\eqref{eta_slab}
and \eqref{alpha_kin_estimate} so that 
$D\simeq-9q\Omega^2h^2/u^2$.
With these assumptions we also obtain the useful relations
$R_\alpha= 3\Omega\tau$ and $R_\Omega= -3(\Omega\tau) qh^2/(\tau u)^2$,
where $\Omega\tau$ is the Coriolis (inverse Rossby) number,
and $h/(\tau u)$ is a dimensionless scale height.

\subsection{Estimating $\xi$}
What is a best estimate for $\xi$ in spiral galaxies?
Observations currently constrain this parameter to only moderate precision,
but $\xi$ is measured to be 
of order unity \citep[e.g][]{Beck07,Beck15a}.
It is important to emphasize, however,
that in the regime of interest,
the large-scale magnetic field is still small $B\ll b$,
whilst observations of nearby galaxies
generally find $B\lesssim b$ \citep{Fletcher10}.

Recent detailed ISM simulations by 
\citet{Kim+Ostriker15} support the value $\xi\sim0.4$ in the fully saturated state. 
\citet{Gent+13b} obtain $\xi\sim0.3$ to $0.6$ as the field approaches saturation,
depending on how $\bfb$ is defined.
Again, here we are more interested in the theoretically predicted
transitory but long-lived regime for which $\bfb$ 
has saturated whilst $\meanv{B}$ is still small.  
This regime is inaccessible in these simulations.
In \citet{Kim+Ostriker15}, the initial large-scale field
has a strength of at least $\sim0.1$ times the value 
corresponding to equipartition with turbulent kinetic energy density.
The small-scale field will thus always be affected by the presence 
of the large-scale field, 
e.g. by tangling of the latter to produce the former.
\citet{Gent+13b} employ a much weaker large-scale seed field,
but the exponential growth rate of $\bfb$ is lower than would be expected,
and even lower than that of $\bfB$ for at least part of the simulation.
The saturation value of $B$ is greater than that of $b$, contrary to observational estimates.
As the authors note, the fluctuation dynamo could be unrealistically weak due to a lack of resolution,
which leads to Reynolds numbers that are too small.
Such simulations have a resolution of a few $\pc$, 
which is much greater than the scales predicted to be most relevant
for fluctuation dynamo growth in the linear regime of $\bfb$.
Thus, obtaining a value of $\xi$ from ISM simulations that is \textit{directly} applicable to our model 
is likely to remain out of reach into the foreseeable future.

One can also turn to results of more generic MHD simulations,
carried out inside a periodic box.
For example,
\citet{Federrath+11} obtain $\xi\sim0.2$--$0.4$
at the end of the exponential growth stage 
in solenoidally forced turbulence simulations 
with Mach number $\mathcal{M}\sim1$,
while the value is smaller by about an order of magnitude
when the turbulence is forced compressively.
These simulations are forced at the scale of the box,
so that the magnetic field can be thought of as small-scale.
They used $\Rey\approx1500$ 
and magnetic Prandtl number $\Pm=\Rm/\Rey\approx2$
(for galaxies $\Pm\gg1$).
Several other simulations of fluctuation dynamo action
have been carried out, 
with saturated values of $\xi$ generally falling somewhere in the range
$\sim$a few $\times0.01$ to $1$, depending on the values of $\mathcal{M}$, $\Pm$, etc. 
\citep[e.g.][see also \citet{Schober+15} for a model]
{Haugen+04,Schekochihin+04,Cho+09,Brandenburg+12a,Bhat+Subramanian13,Federrath+14,Tricco+16}.
Note that $u^2$ in the analytical theory 
characterizes the original turbulence
unaffected by the Lorentz force,
which tends to be somewhat larger than the saturated value of $u^2$,
leading to a somewhat smaller $\xi$.
Taken together, all of these results suggest
that the best estimate of $\xi$ \textit{may} 
be somewhat smaller than $0.4$,
but we consider $0.4$ to be plausible. 
Below we sometimes adopt this value for illustration;
$\xi$ likely also varies within and between galaxies.

How do these estimates
compare with the expected value of $\xi\crit$?
Two realistic parameter sets are explored in 
Models~A$_\xi$ and B$_\xi$ of Table~\ref{tab:models}.
Model~A$_\xi$ has $R_\alpha\approx0.92$ and $R_\Omega\approx-14.1$,
obtained, for example, by setting $q=1$, $u=10\kms$, $h=0.4\kpc$,
and $\tau=10\Myr$, canonical values for the solar neighbourhood
\citep[Ch.~VI of][]{Ruzmaikin+88}.
Model~B$_\xi$ assumes $R_\alpha\approx1.38$ and $R_\Omega\approx-21.1$.
These values are obtained, for instance,
by setting $q=1$, $u=10\kms$, $h=0.2\kpc$, and $\tau=5\Myr$,
as might be appropriate closer to the Galactic centre.
For both models, $\Omega\tau=R_\alpha/3<0.5$. 
This ensures self-consistency with the estimates
of the turbulent transport coefficients,
since those calculations assumed slow rotation,
ignoring quadratic and higher order
terms in $\Omega\tau$.
For these models, our analytic estimate~\eqref{crit} 
yields $\xi\crit\approx0.2$
for Model~A$_\xi$ and $\xi\crit\approx0.4$ for Model~B$_\xi$.
This would imply that the $\alpha$ effect can effectively be reduced 
by a factor of order unity in a real galaxy setting.

\begin{figure}
  \includegraphics[width=1.0\columnwidth,clip=true,trim=0 0  0 0]{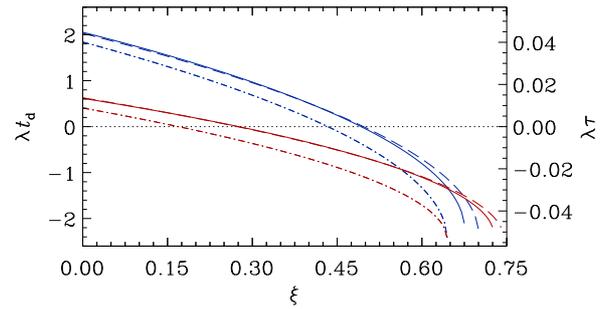}
  %made using Generalized_emf/p.pro
  \caption{Kinematic local growth rate $\lambda$, 
           normalized to the inverse local turbulent diffusion time $t\diff^{-1}$,
           for Models A$_\xi$ (red, bottom set of curves) 
           and B$_\xi$ (blue, top set of curves). 
           Solutions are shown as a function of $\xi=b^2/u^2$,
           sampled in increments of $0.005$ for 
           ($z$-dependent) numerical solutions.
           Full $\alphatilde^2\Omega$ numerical solutions (solid),
           $\alphatilde\Omega$ numerical solutions (dashed),
           and (no-$z$) analytic solutions (dashed-dotted).
           The right axis shows $\lambda$ normalized 
           to the inverse turbulence correlation time $\tau$,
           assuming a locally flat rotation curve $q=1$.
           Curves terminate where solutions become oscillatory (numerical)
           or at $\xi=\xi\f$ (analytical).
           \label{fig:lambdatd_beta}
          }            
\end{figure}

\begin{figure}
  \includegraphics[width=1.0\columnwidth,clip=true,trim=5 0 25 0]{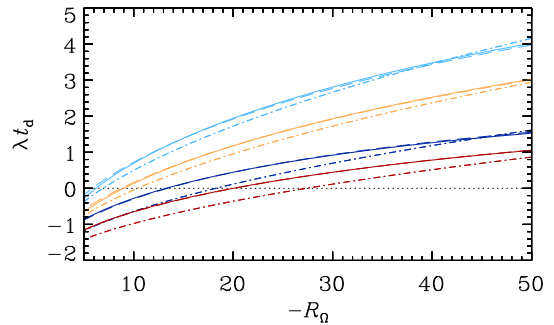}
  %made using Generalized_emf/phoveru.pro
  \caption{Kinematic local growth rate $\lambda$ 
           normalized to the inverse local turbulent diffusion time $t\diff^{-1}$,
           as a function of the Reynolds number $R_\Omega$,
           for Models A$\f$ (orange), A$_{0.4}$ (red, bottom set of curves), 
           B$\f$ (light blue, top set of curves), and B$_{0.4}$ (dark blue).
           Line styles same as Fig.~\ref{fig:lambdatd_beta}.
           \label{fig:lambdatd_ROmega}
          }            
\end{figure}

\subsection{Local growth rate}
\label{sec:growth_rate}
The mean-field induction equation can be solved in the linear regime
under the slab approximation to yield a local (in radius and azimuth) 
mean magnetic field (eigenfunction, which depends on $z$),
and corresponding
exponential growth rate $\lambda$
(eigenvalue, which is independent of $z$).
The growth rate can be estimated as 
\citepalias{Chamandy+14b}
\begin{equation}
  \label{lambdatd}
    \lambda t\diff \sim \frac{\piup^2}{4}\left(\sqrt{\frac{\Dtilde}{D\crit}}-1\right)
    =3\left(\frac{h}{\tau u}\right)^2\lambda\tau,
\end{equation}
where $t\diff=h^2/\eta$ is the turbulent diffusion time,
and we have made use of equation~\eqref{eta_slab} for the rightmost relation.
In Fig.~\ref{fig:lambdatd_beta} 
we illustrate the dependence of $\lambda$ on $\xi$
obtained for Models~A$_\xi$ (red) and B$_\xi$ (blue).
Solid lines show numerical solutions of the full 
$\alphatilde^2\Omega$ dynamo,
whilst dashed lines show solutions that neglect the $\alpha$
and $\delta'$ terms in the $\phi$-component of the mean induction equation
(the $\alphatilde\Omega$ approximation).
Dashed-dotted lines show the analytic solution~\eqref{lambdatd}.
All lines end where the solution becomes oscillatory 
(but retains quadrupole-like symmetry),
or, for the analytic solution, at $\xi=\xi\f$.
The left axis shows $\lambda t\diff$ 
while the right axis shows $\lambda\tau$. 
Right axis labels assume $q=1$.
Note that $\lambda\tau\ll1$; 
thus our implicit assumption
to neglect non-locality in time
is self-consistent, since non-local effects enter
with the factor $(1+\lambda\tau)$ 
in axisymmetric models \citep{Chamandy+13b}.

Numerical solutions yield slightly larger growth rates 
than analytic solutions \citepalias[c.f.][]{Chamandy+14b}, 
but show similar dependence on $\xi$.
For the parameter values chosen,
the $\alphatilde^2$ effect plays only a minor role,
as can be seen by the proximity 
of the solid and dashed curves.
Clearly, the magnetic R\"{a}dler effect is important,
especially for expected solar neighbourhood parameters,
where $\xi\crit\approx0.3$.
The \textit{global} growth rate
is governed by parameter values near the radius 
at which $\lambda$ peaks \citep[e.g.][]{Moss+98b};
values typically assumed 
to be closer to those of Model~B$_\xi$ than Model~A$_\xi$.
Even in this case, the $\delta'$ term is important, 
reducing $\lambda$ by more than half for $\xi=0.4$.

We can explore the parameter space by varying $R_\Omega$.
This is done while holding $\xi$ constant:
either $\xi=0$ in Models~A$\f$ and B$\f$,
or $\xi=0.4$ in Models~A$_{0.4}$ and B$_{0.4}$.
Results are presented in Fig.~\ref{fig:lambdatd_ROmega},
where $\xi=0$ models are shown in lighter colours
(orange and light blue).
Evidently, the effect continues to be important
at other realistic values of $R_\Omega$.

\subsection{Spectral $\tau$ approximation}
\label{sec:spectraltau}
The MTA assumes $\tau$ to be independent of the wavenumber $k$.
Relaxing this assumption 
leads to the spectral $\tau$ approximation.
Under the latter, 
terms proportional to $\tau^2$ get multiplied
by the factor $4/3$, the $\alpha\magn$ term gets modified,
and $\kappa_{ijk}$ gains an extra term 
depending on the spectral index $s$.
Of relevance here, 
equations~\eqref{delta} and \eqref{kappa} become
\citepalias[Appendix~G of][]{Brandenburg+Subramanian05a}
\begin{align}
  \label{delta_spectral}
  &\bm{\delta}= \tfrac{2}{9}\tau^2\bmomega\left(u^2 -b^2\right),\\
  \label{kappa_spectral}
  &\kappa_{ijk}= \tfrac{2}{9}\tau^2
                  \left(\omega_j\delta_{ik} 
                  +\omega_k\delta_{ij}\right)
                  \left[\left(u^2 +\tfrac{7}{5}b^2\right)
                  +\tfrac{2}{5}(s-1)(u^2+b^2)\right].
\end{align}
For $s=1$, we get back MTA, 
whereas $s=5/3$ corresponds to Kolmogorov turbulence.
Using relations~\eqref{delta_spectral} and \eqref{kappa_spectral} 
we obtain, for the slab case, 
$\delta'= -(4/45)\Omega\tau^2[(-1+s)u^2 +(5+s)b^2]$,
which gives $-(8/135)\Omega\tau^2u^2(1+10\xi)$ for $s=5/3$.
That is, we obtain a somewhat stronger effect,
including a (small) part that is independent of $b^2$.
Unfortunately, the $s$-dependent term in $\kappa_{ijk}$
is the one term for which \citetalias{Brandenburg+Subramanian05a}
and \citetalias{Radler+03} do not agree: 
the latter derive an extra factor $-2$.
This leads to the result 
$\delta'= -(8/45)\Omega\tau^2[(1-s)u^2 +(4-s)b^2]$,
or $-(8/135)\Omega\tau^2u^2(-2+7\xi)$ if $s=5/3$.
Thus, in this case $\delta'$ may be close to zero,
or even have opposite sign, 
which would effectively enhance the $\alpha$ effect.
To make progress the discrepancy 
between the two works should be resolved.

\subsection{Nonlinear regime}
\label{sec:nonlinear}
The effect discussed is also relevant in the 
nonlinear (in the \textit{mean} field)
regime of large-scale dynamo action.
It should be kept in mind though that the value of $\xi$ 
will likely be affected by the presence
of the dynamically important large-scale field.
As the effect discussed tends to weaken the dynamo,
one would expect the mean field to saturate
with a strength that is smaller,
given a larger $\xi$.
However, in the dynamical quenching theory,
the $\delta'$ effect also contributes
to the $\Emf\cdot\meanv{B}$ term
in the equation for $\del\alpha\magn/\del t$.
Therefore, the overall effect can be more complicated.
Hence, we leave a study of the nonlinear regime 
for a subsequent work.

\section{Summary and discussion}
\label{sec:discussion}
Starting with an analytical result 
for the mean emf from 
\citetalias{Brandenburg+Subramanian05a}
and \citetalias{Radler+03},
we explore the implications for a generic 
mean-field galactic dynamo model 
in the linear 
regime of the mean field $\meanv{B}$.
We find that the combination of two terms,
involving turbulent transport coefficients
$\delta_i$ and $\kappa_{ijk}$, effectively
suppresses the $\alpha_{ij}$ term for realistic parameter values.
A non-trivial partial suppression from this effect 
depends on the presence of rotation and magnetic fluctuations 
at a level of $(\sim1$ to a few)$\times10\%$
of the turbulent kinetic energy density.
Such conditions are expected to exist 
in the early stages of galaxies,
after saturation of the small-scale dynamo.
This magnetic analogue 
of the R\"{a}dler 
or $\bfOmega\cro\bfJ$ effect thus presents 
a challenge to classical galactic dynamo theory.

Measuring $\Emf$ from appropriate direct numerical simulations
would be valuable to help confirm or falsify the effect discussed,
though this will require significant advances
beyond standard methods \citep{Rheinhardt+Brandenburg10}.
\citet{Brandenburg+08a} and \citet{Brandenburg+12b} measured 
the appropriate coefficients using the test-field method
for statistically homogeneous turbulence with rotation,
with or without density stratification.
However, those studies did not include the feedback
of the magnetic field onto the turbulence from the Lorentz force,
which would be essential. 
Nevertheless, \citet{Brandenburg+12b} obtained 
the expected sign and rough linear dependence on $\Omega$ expected
for $\delta$ and $\kappa$ for the case of homogeneous turbulence with rotation.
In our notation, they found $(k\forc/u)\eta\approx 1/3$,
$[uk\forc/(\eta\Omega)]\delta \approx0.2$, 
and $[uk\forc/(\eta\Omega)]\kappa \approx0.3$
(with $k\forc$ the forcing wavenumber of turbulence).
Putting $b=0$ in equations \eqref{delta_slab} and \eqref{kappa_slab}
with $\eta=(1/3)\tau u^2$ gives $\delta/\eta=\kappa/\eta= (1/2)\Omega\tau$.
Thus, if $\tau\approx1/(k\forc u)$ 
\citep{Brandenburg+Subramanian05b,Brandenburg+Subramanian07},
the analytical results are comparable (larger 
by a factor $\sim2$) for the case $\xi=0$.
This provides rough confirmation of the 
analytical expression of \citetalias{Brandenburg+Subramanian05a}
(assumed throughout the present work)
for the case $\xi\rightarrow0$.
New simulations are needed to test the predictions for 
the case of finite $\xi$.

Our work points to a need
to better constrain the 
ratio of turbulent magnetic 
to kinetic energy density $\xi$.
If, as seems likely, $\xi\approx0.4$
for \textit{any} galaxy known 
to harbour a large-scale magnetic field,
this would suggest a contradiction 
between theory and observation.
It should be emphasized, 
however, that the mean-field dynamo theory used
and the underlying galaxy model are rather approximate,
and results are uncertain by factors of order unity.
Thus, it could simply be the case that $\xi$
is never large enough for the effect to be important.
An alternative remedy to this apparent inconsistency
is a suppression of the fluctuation dynamo,
e.g. by shear \citep{Tobias+Cattaneo13};
but such a scenario seems unlikely \citep{Kolokolov+11,Singh+16}.
In any case, mean-field dynamo models 
sometimes \textit{appeal} to a small-scale dynamo
with saturation strength $\xi\sim1$
to provide sufficient large-scale seeds
of $\sim(0.1$--$1)\nG$ at high redshift 
\citep[e.g.][]{Beck+94,Brandenburg+Urpin98},
in which case a negation of the small-scale dynamo 
would only replace one problem with another.

A related question 
that remains to be addressed is
whether a magnetic shear-current
effect could be important.
Such an effect was explored by \citet{Squire+Bhattacharjee15c},
but not for the high $\Rey$ and $\Rm$ regime relevant for galaxies.
It would therefore be interesting to incorporate shear
into the derivation of the mean emf presented by 
\citetalias{Radler+03} and
\citetalias{Brandenburg+Subramanian05a}.
New contributions to the
small-scale magnetic helicity flux
could also have important effects
\citep{Subramanian+Brandenburg06,Vishniac10,Vishniac+Shapovalov14},
and should be investigated.
More generally, 
our work highlights the
need to clarify the influence 
of small-scale magnetic fluctuations
on the evolution of large-scale magnetic fields
in galaxies and other objects.

\section*{Acknowledgements}
We are grateful to K.~Subramanian for insightful feedback on the manuscript,
and to E.~Blackman, J.~Schober,
J.~Squire and F.~Gent for enlightening comments and discussion.
We also thank the referee for comments that 
led to improvements in the paper.

\footnotesize{
\noindent
\bibliographystyle{mnras}
\bibliography{refs}
}

\label{lastpage}

%-------------------------------------------------------------------------------------------
\end{document}